# Weberite Na$_2$MM'F$_7$ (M, M' = Redox-Active Metal) as Promising Fluoride-Based Sodium-Ion Battery Cathodes

Tenglong Lu[1,2], Sheng Meng[1,2,3]*, Miao Liu[1,3,4]*

[1]*Beijing National Laboratory for Condensed Matter Physics, Institute of Physics, Chinese Academy of Sciences, Beijing 100190, China*
[2]*School of Physical Sciences, University of Chinese Academy of Sciences, Beijing 100190, China*
[3]*Songshan Lake Materials Laboratory, Dongguan, Guangdong 523808, China*
[4]*Center of Materials Science and Optoelectronics Engineering, University of Chinese Academy of Sciences, Beijing 100049, China*

*Corresponding author: smeng@iphy.ac.cn, mliu@iphy.ac.cn

## ABSTRACT

Sodium-ion batteries are a viable alternative to lithium-ion technology due to the plentiful sodium resources. However, certain commercialization challenges, such as low specific energies and poor cycling performance of current Na-ion cathodes, still need to be addressed. To overcome these hurdles, this study explored the potential of a novel class of fluoride-based materials, specifically trigonal-type Na$_2$MM'F$_7$ (M and M' are redox-active metals) belonging to the weberite-type compounds, as promising candidates for Na-ion cathodes. Through a comprehensive assessment utilizing *ab initio* calculations, twelve prospective compounds were identified, demonstrating high thermodynamic stability, large gravimetric capacities (>170 mAh g$^{-1}$), and low net Na-ion migration barriers (<600 meV). Significantly, ten out of the twelve screened compounds exhibit high specific energies exceeding 580 Wh kg$^{-1}$ (approximately equals to the specific energy of LiFePO$_4$), indicating their exceptional electrochemical performance. This study will pave the way for further advancements in fluoride-based electrode materials.

## 1. INTRODUCTION

Electrochemical energy storage, especially in the form of secondary batteries, plays a crucial role in modern life, driving the demand for new materials and improved performance. Lithium-ion battery (LiB) technology is renowned for its high storage capacity and reliable cycling performance. However, the rising cost resulting from high demand and limited abundance of lithium makes it less viable for large storage applications like grid-energy storage.[1] Sodium-ion batteries (SiBs) are viewed as promising lower-cost alternatives to LiBs due to the abundant natural resources of sodium. Among all the critical issues encountered in improving performance of SiBs, the most significant challenge lies in identifying an effective cathode material.

Based on anionic chemistry, the main cathode materials for SiBs can be broadly categorized into three families: layered oxides, polyanionic compounds, and Prussian blue analogues.[1] In comparison to their widely-used Li-ion counterparts, Na-ion layered oxides, such as P2-Na$_{0.6}$CoO$_2$,[2] exhibit lower voltages and sharper voltage profiles. These can be attributed to the higher anode potential of sodium (Na/Na$^+$: -2.71 V and Li/Li$^+$: -3.04 V versus standard hydrogen electrode) and the larger Na-ion radius (116 pm for Na-ion versus 90 pm for Li-ion).[3] These disadvantages result in limited specific energies for Na-ion layered oxides, which in turn restricts their widespread usage as cathode materials for SiBs. In Na-ion polyanionic compounds, the inclusion of polyanionic groups like $(PO_4)^{2-}$ and $(SO_4)^{2-}$ can significantly improve the voltage by reducing the covalent interaction between transition-metal cations and anions, a phenomenon known as the inductive effect.[4] However, enhancing the specific energies of polyanionic compounds necessitates addressing the drawback arising from the higher molar mass of polyanionic groups, which can greatly reduce the gravimetric capacity of the electrode material. As a result, most Na-ion polyanionic cathodes still display constrained specific energies, typically not exceeding 400 Wh kg$^{-1}$.[5]

Fluorine (F) possesses the largest electronegativity, making it the ideal element to initiate the inductive effect in cathode materials. Moreover, compared to polyanionic groups, F$^-$ exhibits a significantly lower molar mass, thereby minimizing its impact on the gravimetric capacity. Due to these factors, researchers have utilized F$^-$ to partially substitute the anionic sites in polyanionic compounds. As a result, the resulting fluorinated polyanionic compounds have emerged as some of the most promising cathode materials for SiBs. Nevertheless, the trade-off between capacity and voltage, although mitigated to some extent, still poses a challenge to the commercialization of these materials. For instance, $Na_3V_2(PO_4)_2F_3$ exhibits an average operating voltage of 3.9 V and a capacity of 128 mAh $g^{-1}$, yielding a specific energy of 499 Wh $kg^{-1}$.[6] Despite its notable improvement in electrochemical activity compared to its NASICON counterpart,[7] it still cannot compete with traditional Li-ion cathode materials (e.g., $LiFePO_4$ delivers a specific energy of 580 Wh $kg^{-1}$). To trigger the inductive effect without compromising capacity, the ideal scenario would entail completely substituting anions or polyanionic groups with $F^-$.

To date, only a limited number of fluoride compounds have been explored as cathodes for SiBs, including $NaFeF_3$, $Na_3FeF_6$, $Na_2MnF_5$, and $NaVF_4$.[8] Unfortunately, these compounds demonstrated poor electrochemical performance. Recently, a novel fluorinated compound, trigonal-type $Na_2Fe_2F_7$, has emerged as a highly promising cathode material.[9] Benefiting from its robust three-dimensional framework, trigonal-type $Na_2Fe_2F_7$ demonstrates outstanding rate capability and structural cyclability. Furthermore, it can deliver a high capacity of 184 mAh $g^{-1}$ (at C/20, 1C = 184 mA $g^{-1}$) along with an average operating voltage of ~3.1 V, making it a promising candidate for electrochemical applications. Its derivative, trigonal-type $Na_2FeTiF_7$, has also been shown to possess an impressive specific energy of ~623 Wh kg$^{-1}$, surpassing other conventional Na-ion cathodes.[10] In contrast, trigonal-type $Na_2MnVF_7$ exhibits only moderate electrochemical performance, with a capacity of ~120 mAh g$^{-1}$.[11] Indeed, these materials hold promise as a new class of fluoride-based cathode candidates, but their development is still predominantly reliant on experimental trial and error within laboratory. Utilizing computational modeling for systematic investigation could significantly expedite the advancement of this kind of materials.

In this work, we performed *ab initio* calculations on a vast range of chemical spaces to evaluate the properties of trigonal-type $Na_2MM'F_7$ (M and M' represent redox-active metals) compounds, focusing on their thermodynamic stability, electrochemical performance, and Na-ion diffusivity. Through our investigation, we successfully identified 12 promising compounds that exhibit remarkable structural stability, high specific energies, and facile Na-ion diffusion pathways, making them suitable candidates for Na-ion cathode materials. Additionally, we conducted a statistical analysis comparing the equilibrium voltages of different redox couples, revealing that species within fluoride chemistry can attain higher redox potentials compared to polyanionic chemistry, such as phosphates. Notably, unlike polyanionic systems, there is no trade-off between voltage and capacity in Na-ion fluorides, making them a category of potentially high-performing electrode materials. It is our hope that this work not only presents several promising cathode compounds but also sets the stage for further advancements in the development of exceptional Na-ion fluoride-based electrode materials.

## 2. METHODS

### 2.1 First-Principles Density Functional Theory Calculations

To obtain the structural energies of various cathode materials, density functional theory (DFT) calculations were conducted using the projector-augmented wave (PAW) method implemented in the Vienna ab initio simulation package (VASP).[12,13] A plane wave energy cutoff of 520 eV and a reciprocal space discretization based on a *k*-point mesh such that $n_{kpoints} \times n_{atoms} > 1000$ were adopted for all calculations. The calculations relied on the Perdew-Burke-Ernzerhof (PBE) generalized gradient approximation (GGA) exchange-correlation functional,[14] with the rotationally averaged Hubbard *U* correction (DFT+*U*) to compensate for the self-interaction error on the specific transition metal atoms.[15] The *U* parameters were obtained from fitting to experimental oxide formation energies,[16] and the detailed values of *U* are in line with the *Materials Project* and *Atomly*.[17–19] All calculations were performed with spin-polarization and initialized in a ferromagnetic high-spin state.

## 2.2 Configurational Enumeration for Disordered Structures

To fully capture all topological distinct Na-vacancy orderings within T-$Na_2MM'F_7$ compounds upon cycling, structural enumeration based on the *EnumLib* package and symmetry analysis based on the *spglib* package were adopted.[20–24] For each configurational enumeration, at least 20 symmetrically distinct structures with minimal electrostatic energies were selected for DFT calculations, and the configuration with the lowest DFT energy was identified as the ground state structure. The electrostatic energy was obtained based on the Ewald summation algorithm.[25]

## 2.3 Phase Diagram Constructions

To evaluate the thermodynamic stability of all the compounds in this work, phase diagrams were constructed with the help of the *Atomly* materials database (containing ~340k compounds). Based on the convex hull method, the stability of any compound was quantified as the value of *energy above hull* ($E_{hull}$), which represents the magnitude of the thermodynamic driving force for the compound to decompose into a set of stable phases in the phase diagram. The python materials genomics (*pymatgen*) open-source library was used to generate all the phase diagrams.[26]

## 2.4 Voltage Profile Calculations

In theory, voltages should be calculated using the Nernst equation:

$$V(x_{Na}) = -\frac{\mu_{Na}^{cathode}(x_{Na}) - \mu_{Na}^{anode}}{nF} \quad (1)$$

where $V(x_{Na})$ is the voltage as a function of the Na-ion concentration ($x_{Na}$), $\mu_{Na}^{cathode}$ and $\mu_{Na}^{anode}$ are the Na chemical potentials in the cathode and anode, $n$ is the charge that is transferred, and $F$ is Faraday constant.

Under several approximations,[27] a simplified Nernst equation is applied in this work:

$$V_{average} = -\frac{E_{Na_{x_2}MM'F7} - E_{Na_{x_1}MM'F7} - (x_2 - x_1)E_{Na}}{(x_2 - x_1)F} \quad (2)$$

where $V_{average}$ is the average voltage of T-Na$_2$MM'F$_7$ in the composition range of $x_1 \leq x \leq x_2$, $E_{Na_xMM'F7}$ and $E_{Na}$ are the internal energies of Na$_x$MM'F$_7$ compound and metallic sodium, respectively.

## 2.5 Migration Barrier Calculations

The migration barriers of Na-ion and vacancy were evaluated using the climbing image nudged elastic band (CI-NEB) method.[28,29] The GGA functional without Hubbard $U$ was adopted to avoid the possible mixing of the diffusion barrier with a charge transfer barrier. Besides, there has been no conclusive evidence showing that GGA+$U$ performs better at predicting cation migration barriers.[30] For all CI-NEB calculations, a reciprocal space discretization of 25 *k*-points per Å$^{-1}$ was applied, and the convergence criteria were set as 5 × 10$^{-5}$ eV for electronic steps and 0.02 eV Å$^{-1}$ for ionic steps.

## 3. RESULTS

### 3.1. Structure Analysis

The electrochemical activity of the trigonal-type Na$_2$MM'F$_7$ compound, hereafter referred to as T-Na$_2$MM'F$_7$, is attributed to its two redox-active metal centers: M and M'. T-Na$_2$MM'F$_7$ belongs to the weberite-type materials and exhibits the space group of *P3$_1$21* with three symmetrically inequivalent Na-ions occupying 6c sites, two-thirds of which are half-occupied (detailed crystal information can be found in Table S1, Supporting Information).[9–11,31] Hence, the as-synthesized T-Na$_2$MM'F$_7$ is a sodium-deficient structure, enabling it to undergo electrochemical cycling between Na$_0$MM'F$_7$ and Na$_2$MM'F$_7$ (as in the case of Na$_2$FeTiF$_7$ compound) or between Na$_1$MM'F$_7$ and Na$_3$MM'F$_7$ (as in the case of Na$_2$Fe$_2$F$_7$ compound), assuming a capacity of 2e$^-$ per formula.[9,10]

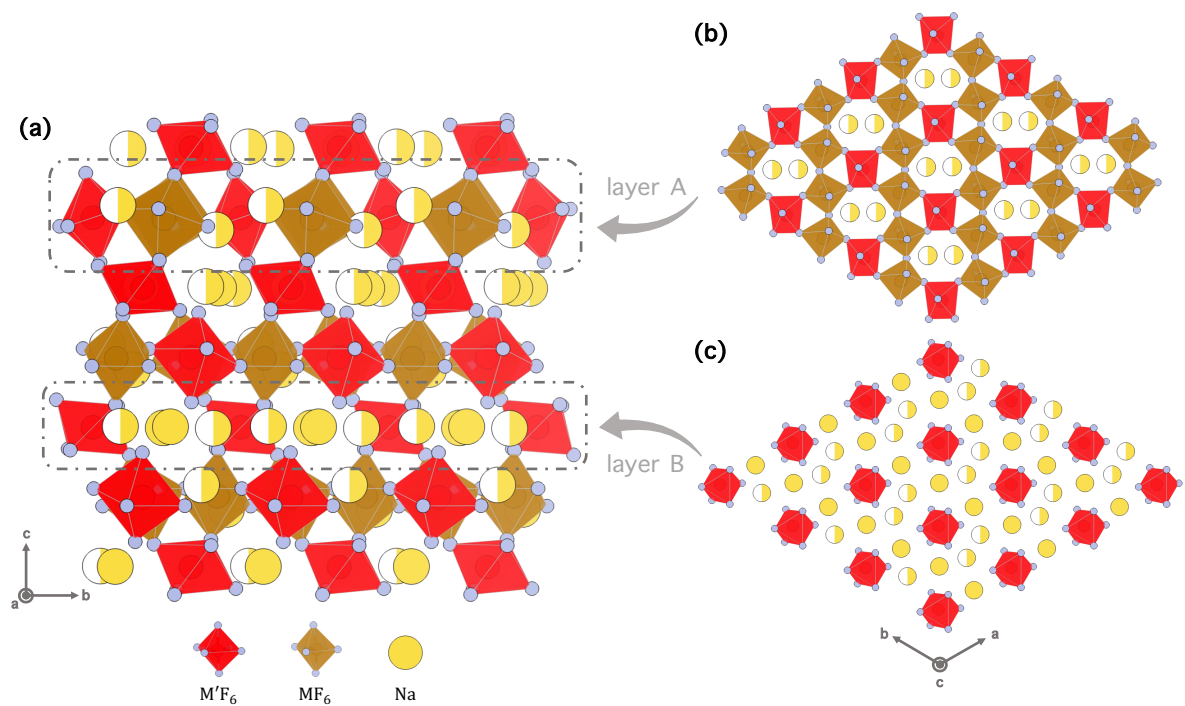

**Figure 1.** (a) Crystal structure of T-Na$_2$MM'F$_7$ (viewed along the [100] direction). The configuration of T-Na$_2$MM'F$_7$ can be regarded as a stacked structure with two stacking components (layer A and B), which are marked by gray dotted boxes. (b) Detailed structure of stacking layer A, viewed along the [001] direction. (c) Detailed structure of stacking layer B, viewed along the [001] direction.

The three-dimensional framework of T-Na$_2$MM'F$_7$, as depicted in Fig. 1a, is composed of corner-sharing redox-active metal octahedrons (MF$_6$ and M'F$_6$). Unlike the distinct separation between the Na layer and the transition-metal layer in the NaCoO$_2$ compound, T-Na$_2$MM'F$_7$ adopts a three-dimensional bulk structure where the distributions of Na and M (M') sites are conjugated. However, the T-Na$_2$MM'F$_7$ compound can still be regarded as a stacked structure with two distinct layers, referred to as layer A (a redox-metal-rich layer) and layer B (a Na-rich layer), as indicated by gray dotted boxes in Fig. 1a. As shown in Fig. 1c, layer B is comprised of isolated M'F$_6$ octahedrons, forming a Na-rich layer that enables the formation of two-dimensional Na-ion diffusion pathways within the *ab*-plane. Layer A is composed of Kagome-like rings consisting four $MF_6$ and two $M'F_6$ octahedrons. As demonstrated in Fig. 1b, the available Na-ion sites are restricted to the center region of each ring, which impedes the formation of in-plane Na-ion diffusion pathways. Nevertheless, layer A can serve as a diffusion channel that connects adjacent layers, allowing for Na-ions to diffuse along the *c*-axis. Hence, the T-$Na_2MM'F_7$ compound demonstrates a substantial three-dimensional Na-ion diffusion network, which can play a curcial role in maintaining its structural stability during cell cycling. Furthermore, the presence of a corner-sharing framework has been shown to greatly enhance metal-ion diffusion, as demonstrated by a recent study on solid-state electrolytes such as $LiTa_2PO_8$ and $LiGa(SeO_3)_2$.[32] Consequently, it is expected that T-$Na_2MM'F_7$ compounds will exhibit outstanding structural cyclability and rate capability, suggesting their potential as a class of exceptional cathode materials for SiBs.

## 3.2. Structure Generation

To systematically evaluate the properties of T-$Na_2MM'F_7$, we utilized a number of redox-active species, including Ti, V, Cr, Mn, Fe, Co, Ni, Cu, Nb, Mo, Sn, Sb, and Bi, to replace the M and M' sites within the structure. Taking into account the evolution of oxidation states during cell cycling, we excluded Nb, Mo, Sb, and Bi from occupying the M sites (+2 oxidation state, Wyckoff position: 6c), and Cu from occupying the M' sites (+3 oxidation state, Wyckoff position: 3a and 3b).[33] For each $Na_2MM'F_7$ compound, its ordered structure and three (de)sodiated counterparts, namely $Na_0MM'F_7$, $Na_1MM'F_7$ and $Na_3MM'F_7$, were generated based on a structural enumeration method (see Methods). Finally, we obtained a total of 108 M-M' metal pairs and 432 crystal configurations.

### 3.3. Thermodynamic Stability

Next, the energies of these 432 compounds were determined by employing *ab initio* calculations. The thermodynamic stability of each Na-M-M'-F compound, quantified as the energy above the convex hull ($E_{hull}$), can be derived based on the calculation results along with the extensive data in the *Atomly* materials database. Essentially, the $E_{hull}$ of a compound represents the thermodynamic driving force for the compound to decompose into the most stable phases located on the convex hull. Materials with $E_{hull} = 0$ are regarded as thermodynamically stable phases, whereas materials with $E_{hull} > 0$ are considered as metastable (or unstable) phases.

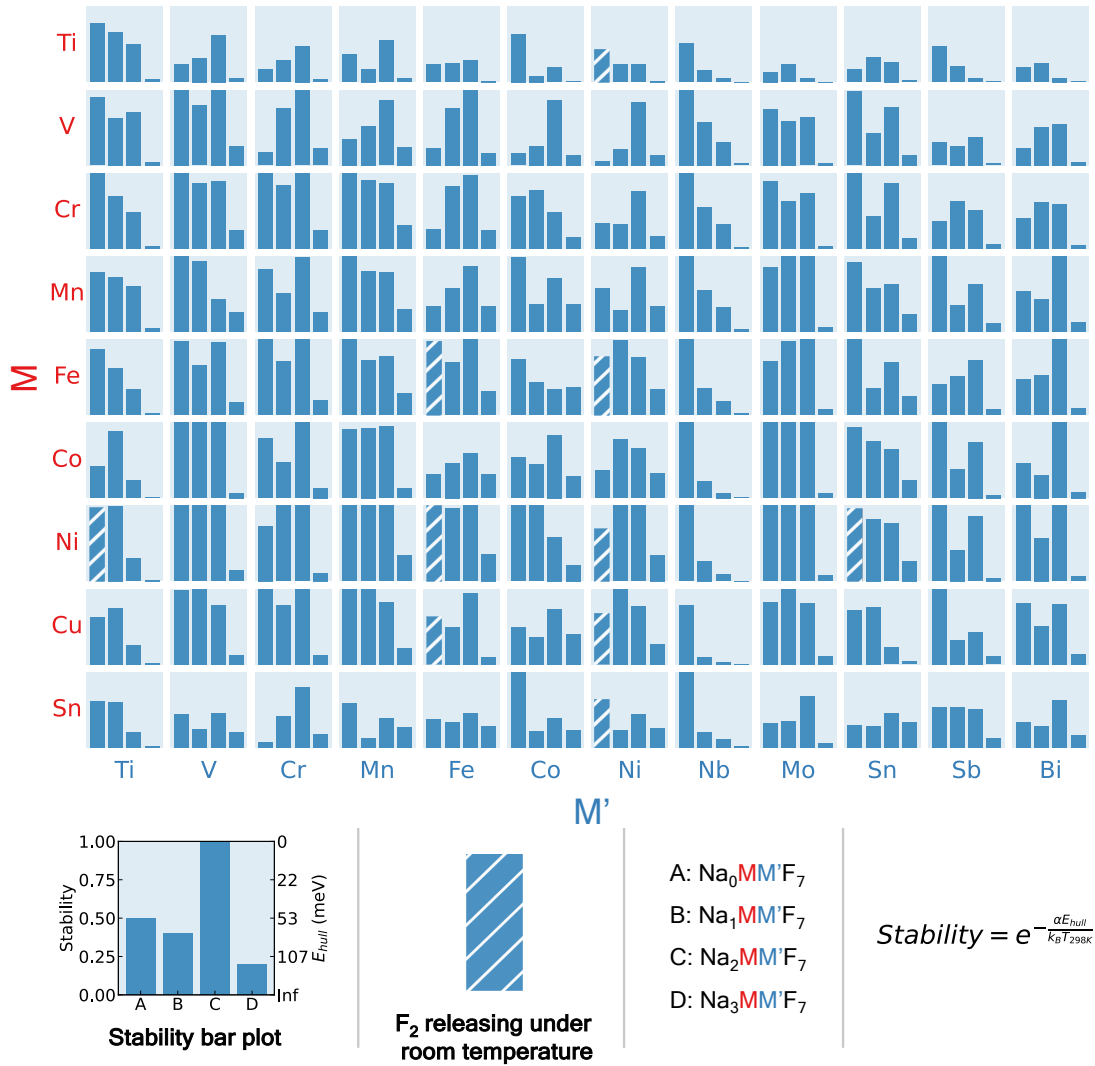

**Figure 2. Thermodynamic stability map for Na$_x$MM'F$_7$ (x = 0, 1, 2, 3) compounds.** For each M-M' metal pair, a bar plot is utilized to illustrate the thermodynamic stability of Na$_0$MM'F$_7$, NaMM'F$_7$, Na$_2$MM'F$_7$, and Na$_3$MM'F$_7$, in which the height of each bar represents the value of stability. The stability of a compound is quantified by converting the $E_{hull}$ to a value through the utilization of the Boltzmann factor, as shown in the lower right panel (where α, an adjustment factor, is set to 1/3 in this work). Bars with stripes represent compounds undergoing F$_2$ releasing under room temperature.

Figure 2 visualizes the thermodynamic stability of 432 compounds across 108 M-M' metal pairs [M in (Ti, V, Cr, Mn, Fe, Co, Ni, Cu, Sn), and M' in (Ti, V, Cr, Mn, Fe, Co, Ni, Nb, Mo, Sn, Sb, Bi)] at different Na-ion concentrations, encompassing stoichiometries of Na$_0$MM'F$_7$, Na$_1$MM'F$_7$,

Na$_2$MM'F$_7$, and Na$_3$MM'F$_7$. To better illustrate the thermodynamic stability, in Fig. 2, we convert the $E_{hull}$ value into a stability index, which quantifies the likelihood of a compound's existence based on the Boltzmann factor as:

$$Stability = e^{-\frac{\alpha E_{hull}}{k_B T_{298K}}} \qquad (3)$$

Hence, stability = 1 ($E_{hull}$ = 0 meV/atom) represents a stable compound, whereas stability < 0.25 ($E_{hull}$ > 107 meV/atom) indicates that the compound is highly unstable and less likely to exist. Additionally, compounds that are prone to decompose into the F$_2$ gas phase at room temperature are identified by bars with diagonal stripes (see details in Section II, Supporting Information).

Figure 2 serves as a stability map for exploring synthesizable compounds that demonstrate viable thermodynamic stability. Certain trends can be observed by examining this map. For instance, it can be found that the presence of Ni in M site (Wyckoff position: 6c) and Ti, V, or Mn in M' (Wyckoff position: 3a and 3b) site tends to enhance the stability of the phase. Conversely, when Ti, V, or Mn occupy the M site and Ni is present in the M' site, the structure tends to exhibit reduced stability. The denser Na-ion sublattices in Na$_3$MM'F$_7$ compounds result in stronger electrostatic repulsion, leading to their moderate to infernal thermodynamic stability. This highlights the general challenge in reducing both the M and M' species to the +2 oxidation states. Given that $Na_0MM'F_7$ compounds are typically stable phases, it is less likely to directly discharge these compounds to $Na_3MM'F_7$ phases, as two-electron transfer rarely happens for redox-active metals. In Fig. 2, the differentiation of the M and M' sites provides valuable insights into the specific site preferences of redox-active metals. For example, $Na_2NiVF_7$ ($E_{hull}$ = 0 meV/atom, stability = 1) is more stable than $Na_2VNiF_7$ ($E_{hull}$ = 14 meV/atom, stability = 0.83), indicating that Ni prefers the M site and V prefers the M' site. Hence, the Na-Ni-V-F compound tends to adopt the more stable $Na_2NiVF_7$ phase during the synthesis. Compounds containing Fe and Ni species, when fully charged to $Na_0MM'F_7$, are vulnerable to the release of $F_2$ gas, indicating poor thermal stability.[34,35] A previous work demonstrated that fluorides show a large fraction of metastable phases. And the unusual metastability of fluorides can be attributed to the strong metal-fluorine bonds in the solid state.[36] Hence, we adopted an $E_{hull}$ threshold of 107 meV/atom as the metastable limit in this work. $Na_2Fe_2F_7$ ($E_{hull}$ = 0 meV/atom), $Na_2FeTiF_7$ ($E_{hull}$ = 85 meV/atom), and $Na_2MnVF_7$ ($E_{hull}$ = 65 meV/atom) have been synthesized experimentally,[9–11] implying the accuracy of our stability evaluation.

### 3.4. Electrochemical Performance

Based on the calculation results of the 108 M-M' metal pairs, the electrochemical performance of T-Na$_2$MM'F$_7$ compounds as Na-ion cathode materials can be derived according to the Nernst equation. Figure 3 exhibits the voltage profiles (step-wise lines within each box) and the specific energies (shades of color for each box) for all the Na-M-M'-F compounds. To ensure compatibility with the typical electrochemical stability window of Na-ion electrolytes,[37] the viable Na-ion cycling range is defined to be within the voltage range of 1.2-4.5 V. Na$_2$CoCrF$_7$, Na$_2$NiCrF$_7$, Na$_2$VCoF$_7$, Na$_2$VNiF$_7$, Na$_2$CrNiF$_7$, Na$_2$CoBiF$_7$, and Na$_2$NiBiF$_7$ compounds do not present electrochemical activity as their voltage profiles fall outside the viable voltage range (marked with diagonal stripes in Fig. 3). Therefore, these compounds are excluded from further screening.

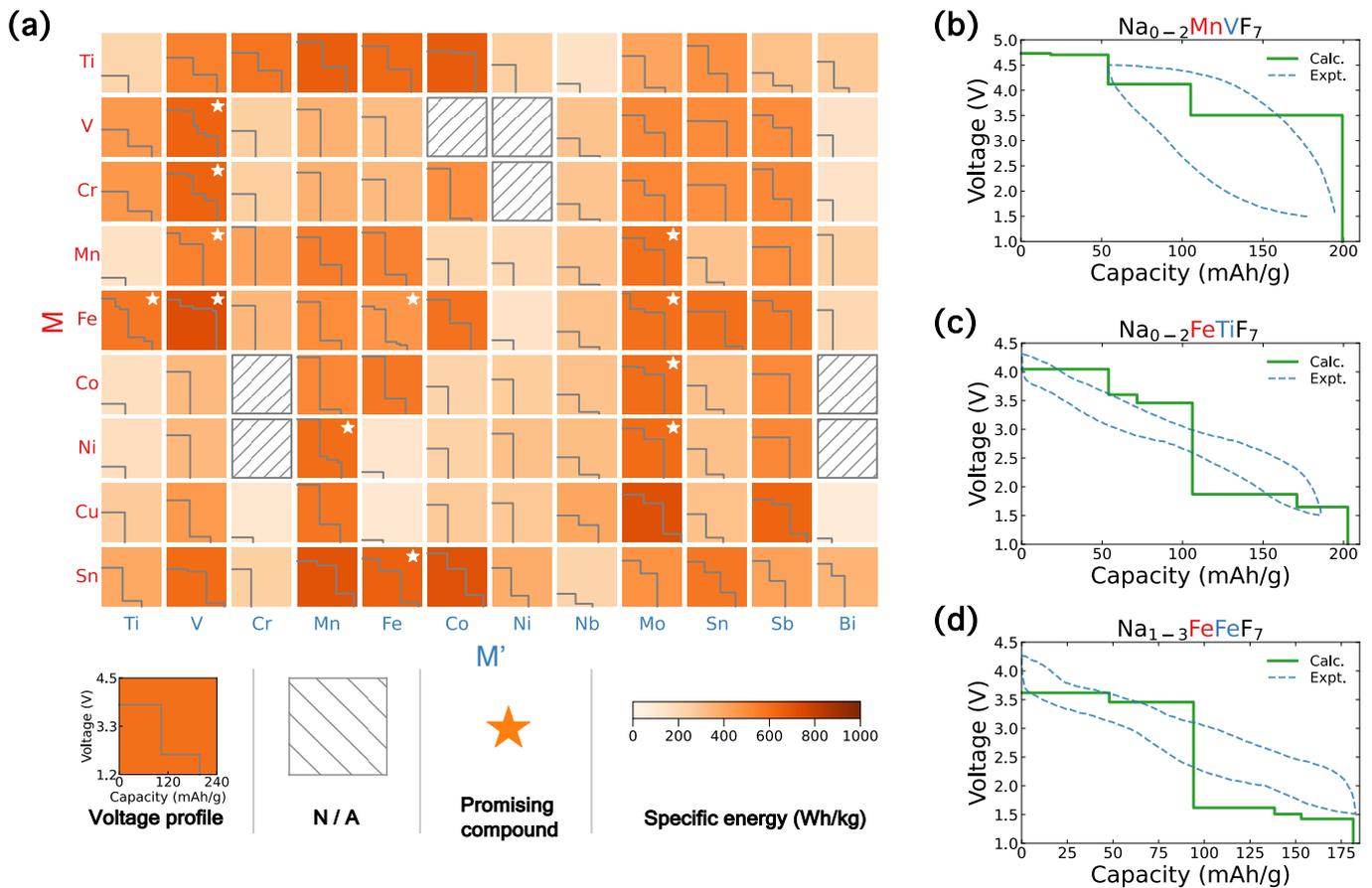

**Figure 3.** (a) Electrochemical performance of 108 Na-M-M'-F chemical systems. Voltage profiles are shown as step-wise lines in each box. Color represents specific energy. Compounds with starts stand for promising cathode candidates. Theoretical and experimental voltage profiles of (b) $Na_{0-2}MnVF_7$, (c) $Na_{0-2}FeTiF_7$, and (d) $Na_{1-3}Fe_2F_7$.

Based on the calculation results shown in Fig. 2 and Fig. 3, several promising cathode candidates (marked with stars in Fig. 3), demonstrating viable synthesizability, feasible voltage profiles, and high specific energies, have been identified, including $Na_2V_2F_7$, $Na_2CrVF_7$, $Na_2MnVF_7$, $Na_2FeVF_7$, $Na_2FeTiF_7$, $Na_2NiMnF_7$, $Na_2Fe_2F_7$, $Na_2SnFeF_7$, $Na_2MnMoF_7$, $Na_2FeMoF_7$, $Na_2CoMoF_7$, and $Na_2NiMoF_7$. Note that for all these compounds, their detailed theoretical voltage profiles have been fine-calculated by incorporating a large number of Na-ion concentrations between the charged and discharged states. For each Na-ion concentrarion, multiple Na-vacancy orderings are calculated to accurately capture the ground states and ensure the correct voltage versus capacity curve. The detailed pseudo-binary phase diagrams of these 12 cathode candidates are illustrated in Fig. S1, Supporting Information.

$Na_2MnVF_7$, $Na_2FeTiF_7$, and $Na_2Fe_2F_7$ are three promising cathode candidates that have been experimentally synthesized in previous studies.[9–11] As shown in Fig. 3b-3d, their theoretical voltage profiles are in good agreement with the experimental observations, implying the accuracy of our calculation results. For $Na_2MnVF_7$ compound, there appears to be a voltage plateau at around 4.7 V, which corresponds to the compositions between $Na_0MnVF_7$ and $Na_{0.5}MnVF_7$. However, this equilibrium voltage has not been experimentally observed due to the operating voltage limitations imposed by the electrolytes, which restrict the cutoff potential to 4.5 V.[11] If the electrochemical window of Na-ion electrolytes can be expanded to about 5 V, our theoretical data indicates that $Na_2MnVF_7$ has the potential to deliver a specific energy of 796 Wh kg$^{-1}$, which is comparable to that of the state-of-the-art Li(Ni, Mn,Co)O$_2$ (i.e., NMC) cathode.

**3.5. Na-ion Diffusivity**

The diffusivity of the mobile ion (e.g., Li-ion, Na-ion, Mg-ion) largely controls the aspects of cathode kinetic performance, such as rate capability, polarization, and capacity retention. This effect is especially pronounced in Na-ion cathodes because of the relatively larger size of Na-ions compared to other mobile ions. Consequently, we conducted an evaluation of Na-ion diffusion properties for all 12 cathode candidates.

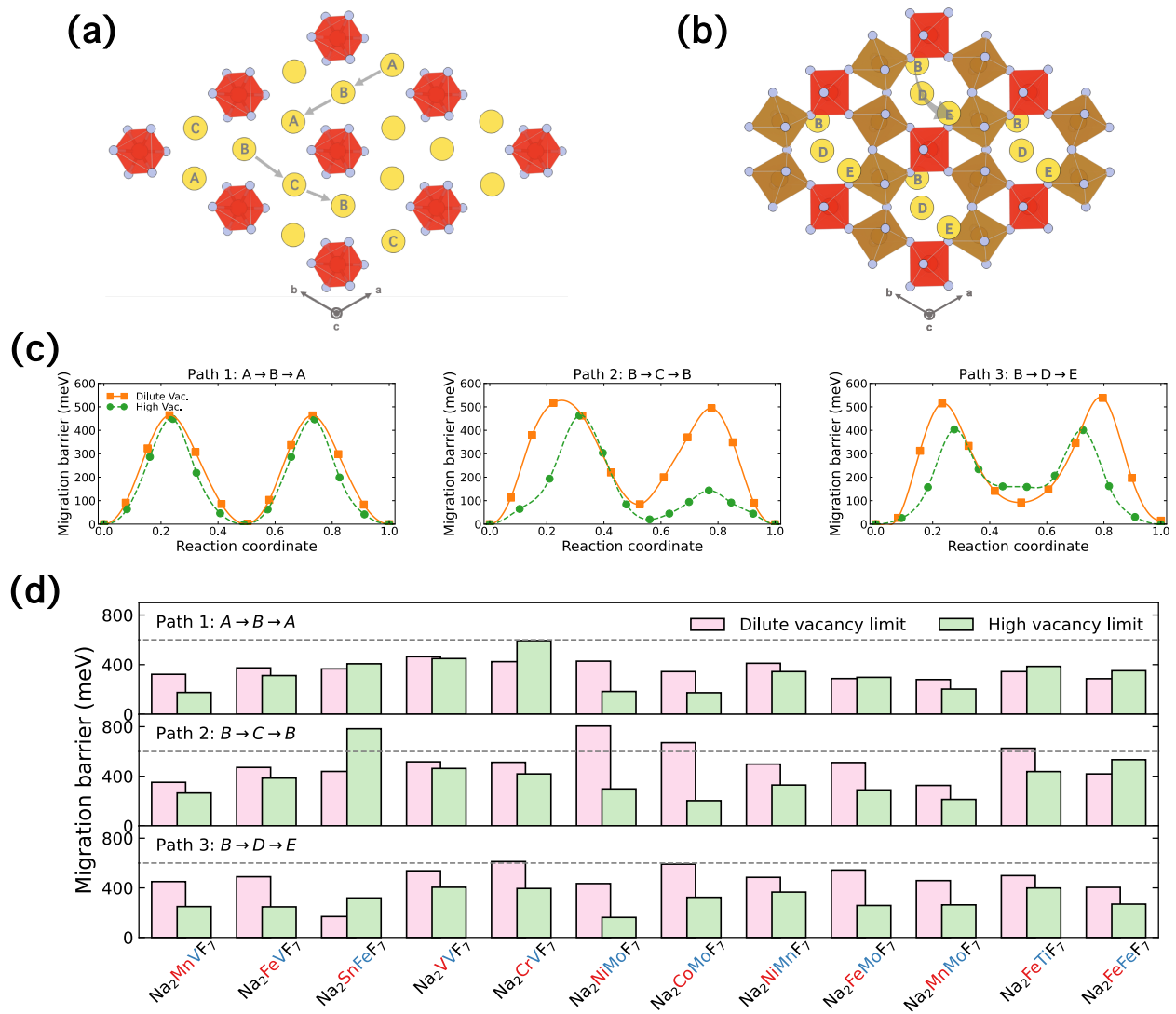

**Figure 4.** Schematic diagram of Na-ion migration pathways for a T-Na$_2$MM'F$_7$ compound (a) within *ab*-plane and (b) along *c*-axis. Three distinct pathways are identified: along sites A-B-A (path 1), along sites B-C-B (path 2), and along sites B-D-E (path 3). (c) Detailed Na-ion migration barriers of the Na$_2$V$_2$F$_7$ compound along reaction coordinates for three diffusion paths under both dilute (orange solid lines) and high (green dashed lines) vacancy limits. (d) Bar plot of migration barriers for all the 12 candidate compounds considered in this work. The dashed lines in the plot indicate a migration barrier threshold of 600 meV.

Figure 4a and 4b give schematic representations of the Na-ion diffusion pathways for T-Na$_2$MM'F$_7$ compounds. Three distinct diffusion paths are identified: path 1 (along sites A-B-A), path 2 (along sites B-C-B), and path 3 (along sites B-D-E). Path 1 and path 2 are within the *ab*-plane (i.e., layer

B in Fig. 1c), forming an in-plane two-dimensional percolation network. Path 3 is along the *c*-axis, connecting the Na-ion sites within adjacent layers, thereby creating a three-dimensional diffusion network for Na-ions. It should be noted that Na-ions can percolate through the structure with any of the three abovementioned pathways.

Figure 4c illustrates the detailed Na-ion migration barriers of $Na_2V_2F_7$ compound along the reaction coordinates for all three diffusion pathways. As the mobile-ion diffusivity depends on the cathode states of charge, we calculated the migration barriers at both dilute (corresponding to the discharged state) and high (corresponding to the charged state) vacancy limits. As shown in Fig. 4c, $Na_2V_2F_7$ exhibits Na-ion migration barriers of ~500 meV for the dilute vacancy limit and ~440 meV for the high vacancy limit. For comparison, $Na_3V_2(PO_4)_2F_3$, a highly promising vanadium-based Na-ion cathode, shows net Na-ion barriers of ~600 and ~300 meV in the dilute and high vacancy limits, respectively.[38] Hence, it is anticipated that $Na_2V_2F_7$ will demonstrate at least comparable kinetic performance to $Na_3V_2(PO_4)_2F_3$.

Figure 4d and Table 1 present Na-ion migration barrier information for all 12 candidate compounds. Their detailed migration profiles can be found in Fig. S2-S12, Supporting Information. For both Na-ion vacancy limits, the results indicate that the Na-ion migration barriers for all the 12 candidate compounds are mostly below a threshold value of 600 meV, suggesting their favorable Na-ion mobilities and promising kinetic performance as cathode materials. Besides, the presence of a three-dimensional diffusion network in T-$Na_2MM'F_7$ compounds enhances their robustness as Na-ion conductors, as there are always facile routes for Na-ions to transport even when one of the pathways becomes sluggish. For instance, although the Na-ion migration barrier along path 2 in $Na_2FeTiF_7$ exceeds 600 meV at the dilute vacancy limit, the barrier along path 1 is only 344 meV, which still ensures the overall facile Na-ion transport, as evidenced by its outstanding capacity retention of 71% after 600 cycles at 189 mA $g^{-1}$.[10] Compared to $Na_2FeTiF_7$, as demonstrated in Table 1, $Na_2Fe_2F_7$ exhibits substantially lower Na-ion migration barriers, which aligns with its significantly improved capacity retention of 88% after 1000 cycles at 368 mA $g^{-1}$.[9] Moreover, our prediction suggests that $Na_2MnMoF_7$ compound, with low enough Na-ion migration barriers along any of the three pathways in both vacancy limits, is expected to display superior kinetic performance, potentially even surpassing $Na_2Fe_2F_7$.

**Table 1. Calculated Na-ion migration barriers of promising candidates for Na-ion cathodes.** For experimentally investigated $Na_2MnVF_7$, $Na_2FeTiF_7$ and $Na_2Fe_2F_7$ compounds, their capacity retentions after several hundred cycles were presented.

| Formula | Path 1 (meV) | | Path 2 (meV) | | Path 3 (meV) | | Cycle Number | Capacity Retention |
|---|---|---|---|---|---|---|---|---|
| | Dilute Vac. | High Vac. | Dilute Vac. | High Vac. | Dilute Vac. | High Vac. | | |
| $Na_2MnVF_7$ | 322 | 176 | 351 | 265 | 451 | 249 | 200 @ 100 mA g$^{-1}$[11] | 93%[11] |
| $Na_2FeVF_7$ | 374 | 312 | 471 | 385 | 490 | 247 | — | — |
| $Na_2SnFeF_7$ | 366 | 407 | 439 | 783 | 169 | 319 | — | — |
| $Na_2V_2F_7$ | 464 | 449 | 517 | 463 | 538 | 405 | — | — |
| $Na_2CrVF_7$ | 424 | 594 | 513 | 419 | 612 | 395 | — | — |
| $Na_2NiMoF_7$ | 428 | 184 | 805 | 298 | 435 | 162 | — | — |
| $Na_2CoMoF_7$ | 344 | 174 | 671 | 203 | 591 | 324 | — | — |
| $Na_2NiMnF_7$ | 411 | 344 | 498 | 329 | 485 | 366 | — | — |
| $Na_2FeMoF_7$ | 288 | 298 | 511 | 290 | 544 | 258 | — | — |
| $Na_2MnMoF_7$ | 280 | 203 | 326 | 213 | 459 | 263 | — | — |
| $Na_2FeTiF_7$ | 344 | 386 | 626 | 438 | 499 | 399 | 600 @ 189 mA g$^{-1}$[10] | 71%[10] |
| $Na_2Fe_2F_7$ | 287 | 351 | 419 | 534 | 404 | 269 | 1000 @ 368 mA g$^{-1}$[9] | 88%[9] |

### 3.6. Overall Performance

In summary, Figure 5 presents a radar chart that comprehensively depicts the overall performance of the 12 promising cathode candidates, considering factors such as thermodynamic stability, gravimetric capacity, average voltage, specific energy, energy density, and migration barrier. For comparison, the overall performance of P2-$Na_{2/3}CoO_2$ and $Na_2Fe_2(SO_4)_3$,[39,40] which serve as representative examples of Na-ion layered and polyanionic oxides, are also presented in the chart.

It is evident that all 12 candidate compounds have the potential to serve as Na-ion cathodes and outperform certain known SiB cathode materials, such as P2-$Na_{2/3}CoO_2$ and $Na_2Fe_2(SO_4)_3$. In most cases, the Na-M-M'-F compound undergoes a 2e⁻ redox reaction per formula, resulting in a gravimetric capacity of at least 170 mAh g$^{-1}$. For vanadium-based systems, their gravimetric capacities can reach as high as approximately 200 mAh g$^{-1}$. Furthermore, when the Sn species is involved, the compound can undergo a 3e⁻ redox reaction per formula, leading to a significant increase in the gravimetric capacity of up to about 230 mAh g$^{-1}$.

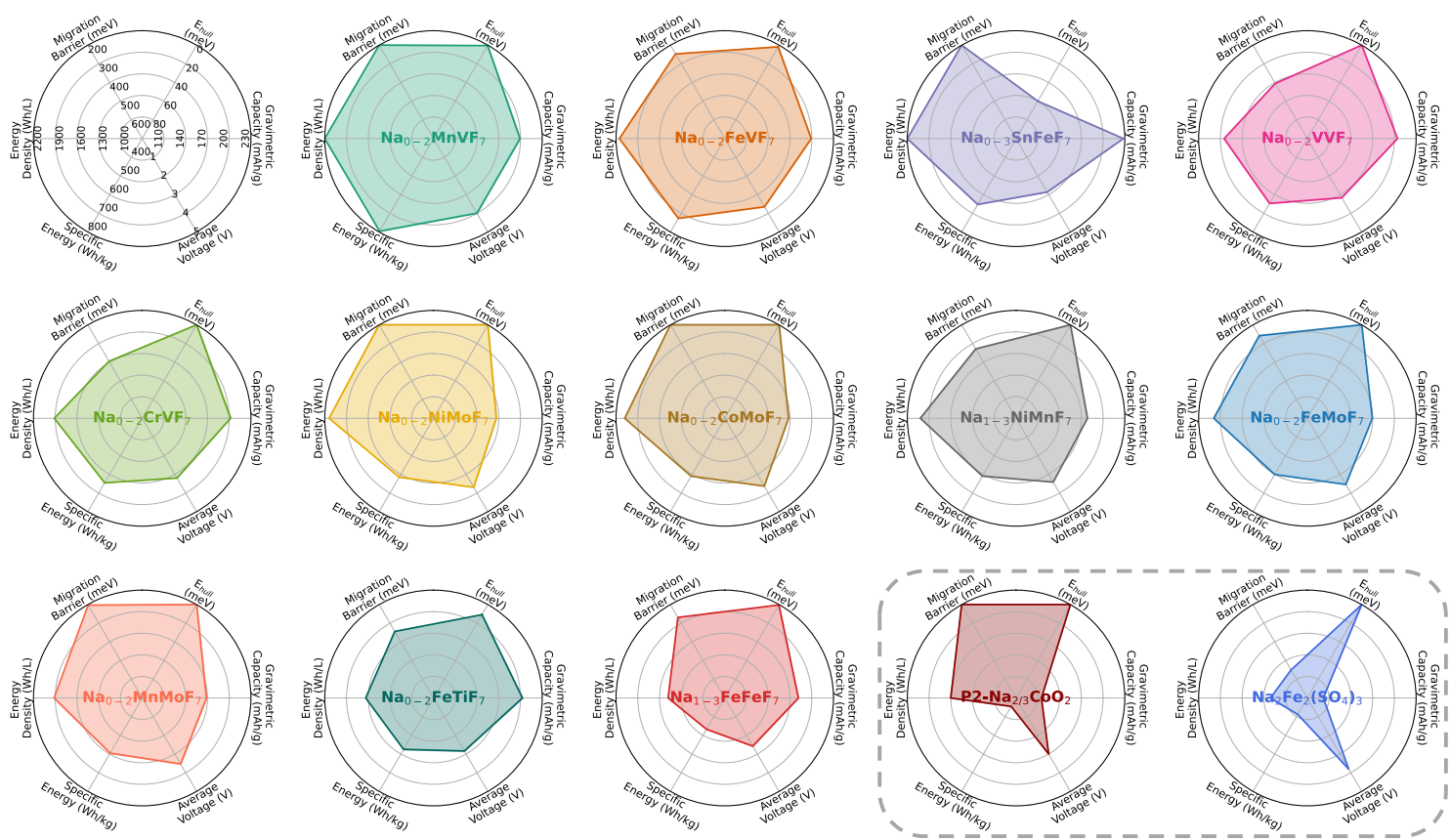

**Figure 5.** Radar chart of the overall performances for the 12 candidate compounds. The performances of layered P2-Na$_{2/3}$CoO$_2$ and polyanionic Na$_2$Fe$_2$(SO$_4$)$_3$ cathodes are also presented for comparison.

## 4. DISCUSSION

With the maximum electronegativity, fluorine is an ideal choice to trigger the inductive effect within the electrode context without compromising the capacity. This characteristic is crucial for Na-ion cathodes to achieve the electrochemical performance of current commercialized Li-ion electrodes. While this study focused on a specific structural system, it is worth noting that the equilibrium voltage delivered by a redox couple is an inherent property of the species, which enables us to derive the general redox potentials for species in fluoride chemistry.

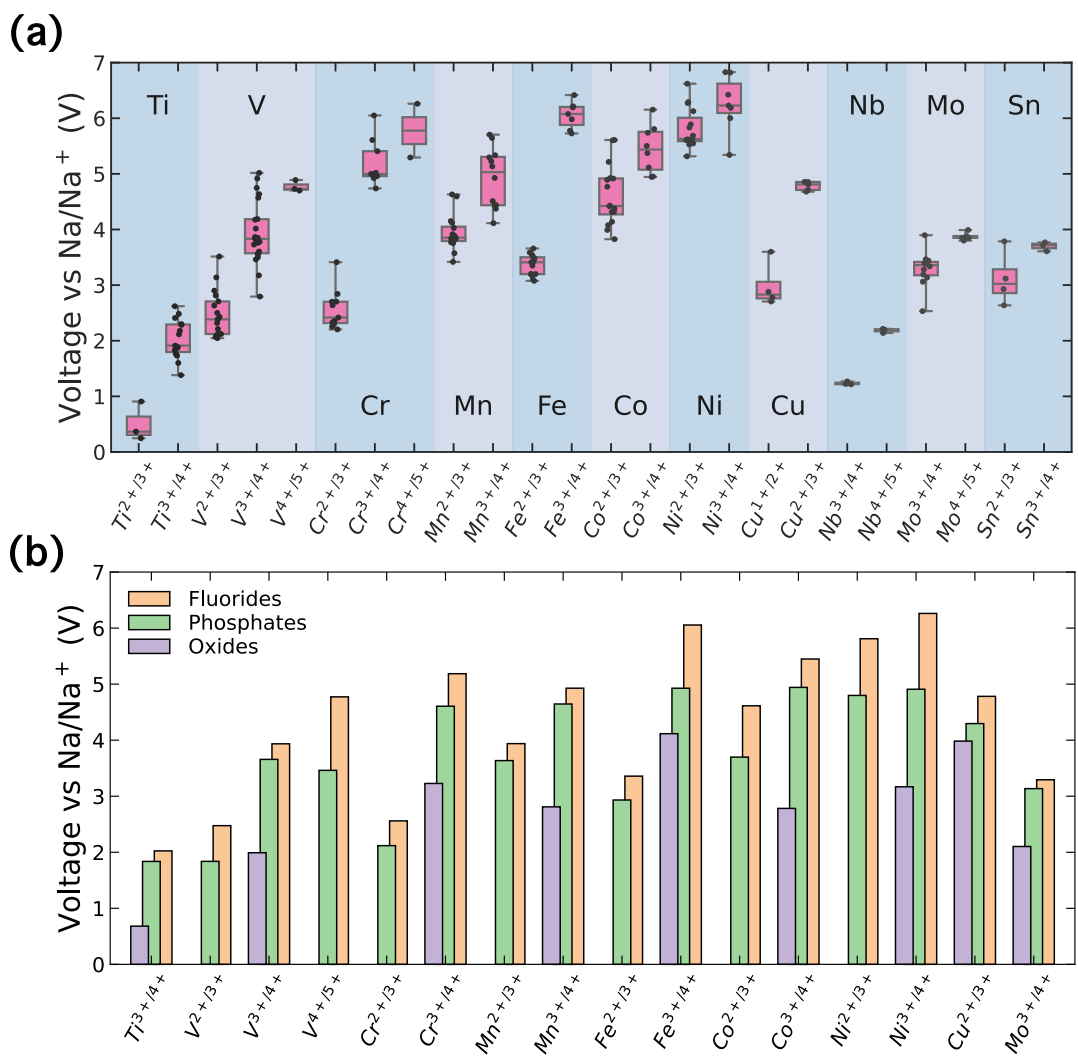

**Figure 6.** (a) Equilibrium voltages delivered by different redox couples in fluoride chemistry. Each black point stands for a compound. For each redox couple, the pink box represents the quartiles of the data set while the whiskers extend to show the rest of the distribution. (b) Average equilibrium voltages of redox couples in different anionic chemistries, including oxides (purple), phosphates (green), and fluorides (yellow). Note that in panel b, only the chemistries that are available within an internal electrode database were presented.

Figure 6a presents the equilibrium voltages of different redox couples in Na-ion fluorides, considering only compounds that are reasonably stable in their sodiated states ($E_{hull}$ < 100 meV/atom). For comparison, Figure 6b showcases the average voltages of various redox couples in different anionic chemistries, including oxides, phosphates, and fluorides. From Fig.

6a, it is evident that the average voltage for each element increases with the oxidation state (e.g., the average voltage for $Fe^{3+}/Fe^{4+}$ is higher than that for $Fe^{2+}/Fe^{3+}$), which is consistent with the trend observed in oxides. Additionally, as depicted in Fig. 6b, species in fluorides deliver higher redox voltages than those in phosphates, indicating a more potent inductive effect in fluoride chemistry. For instance, triphylite $NaFePO_4$ exhibits an average $Fe^{2+}/Fe^{3+}$ redox potential of 2.8 V versus $Na/Na^+$ in experiments (agrees well with our prediction of 2.9 V),[41] which is 0.6 V lower than the redox potential of $LiFePO_4$.[42] This voltage drop leads to a reduction in the specific energy of $NaFePO_4$, which in turn hinders its practical application. On the other hand, we predict that the $Fe^{2+}/Fe^{3+}$ redox couple in fluorides can demonstrate a high voltage of 3.36 V, which is comparable to that of $LiFePO_4$. Indeed, the experimental results have demonstrated that the inclusion of $F^-$ in the system enables the fluorophosphate $Na_2FePO_4F$ to achieve a higher redox plateau at 3.0 V.[43]

The voltage profiles of cathode materials are largely determined by the equilibrium voltages delivered by redox couples. As mentioned in Section 3.4, the voltage plateau at 4.7 V in the $Na_2MnVF_7$ curve leads to a deviation between theory and experiment. We found that this 4.7 V potential is related to the $V^{4+}/V^{5+}$ redox couple, implying that half of the Mn ions in the compound are still in the +2 oxidation state at the fully charged state

($Na_0MnVF_7$). This is an abnormal phenomenon, considering the $Mn^{2+}/Mn^{3+}$ potential (~4 V) is lower than that of the $V^{4+}/V^{5+}$ (~4.7 V), but can be attributed to a strain effect associated with the strong Jahn-Teller distortion of $Mn^{3+}$ ions (details in Section V, Supporting Information). Since there are two redox-active metal sites per formula in T-$Na_2MM'F_7$ compounds, redox competition between M and M' is expected to occur. For instance, $Na_2NiMoF_7$ (Fig. 6f), $Na_2CoMoF_7$ (Fig. 6g), and $Na_2MnMoF_7$ (Fig. 6j) all display very similar voltage profiles, which should be attributed to the same redox reaction originating from the common Mo species. As shown in Fig. 6a, redox potentials of Mo couples ($Mo^{3+}/Mo^{4+}$ and $Mo^{4+}/Mo^{5+}$) are all lower than those of Ni, Co, and Mn couples ($M^{2+}/M^{3+}$, M = Ni, Co, Mn), implying that the redox activities of Ni, Co, and Mn species are actually suppressed throughout the electrochemical cycling.

Fluorides generally show poor electronic conductivity due to localized electrons caused by strong ionic bonds. Experimentally, polarization was observed in the voltage curves of $Na_2MnVF_7$ (Fig. 3b), $Na_2FeTiF_7$ (Fig. 3c), and $Na_2Fe_2F_7$ (Fig. 3d). Considering the facile Na-ion diffusion (see Section 3.5), the voltage polarization should be attributed to the low electronic conduction in these compounds. We used the Kohn-Sham band gap from DFT calculations as a rough indicator for electronic conductivity.

It should be noted that our DFT calculations were at the GGA+$U$ level, which would severely underestimate band gaps. Nonetheless, the band gap values we obtained are still valuable for capturing the general trends. As shown in Fig. S15, Supporting Information, the band gap of $Na_2MnVF_7$ is about 2 eV higher than those of $Na_2FeTiF_7$ and $Na_2Fe_2F_7$, which is consistent with the larger polarization observed in the voltage curve of $Na_2MnVF_7$ (Fig. 3b). Besides, $Na_2NiMoF_7$, $Na_2FeMoF_7$, and $Na_2MnMoF_7$ may exhibit even more pronounced polarization than $Na_2MnVF_7$. For the complete chemical spaces considered in this work, a band gap heatmap is presented in Fig. S16, Supporting Information. To address the issue of poor electronic conduction in T-$Na_2MM'F_7$ compounds, the methods already developed in $LiFePO_4$, such as nanosizing processes and homogenous particle coatings of conductive materials,[42] should also be valid for this class of materials.

## 5. CONCLUSIONS

Anion chemistry plays a critical role in enhancing the electrochemical performance of Na-ion cathodes due to inherent limitations of SiBs, such as the high Na/$Na^+$ redox potential and the large Na-ion radius. This study explored the utilization of fluorine, known for its maximum electronegativity, to trigger the inductive effect without compromising the capacity. We conducted a systematic evaluation of the thermodynamic stability, electrochemical performance, and Na-ion diffusion mechanism for a novel class of fluoride-based cathode materials known as T-$Na_2MM'F_7$, which belongs to the weberite-type materials. Through the screening of 108 M-M' metal pairs, we identified 12 promising Na-ion cathode materials that exhibit remarkable thermodynamic stability at any states of charge, as well as high specific energies and facile Na-ion diffusion networks, including $Na_2MnVF_7$, $Na_2FeVF_7$, $Na_2SnFeF_7$, $Na_2V_2F_7$, $Na_2CrVF_7$, $Na_2NiMoF_7$, $Na_2CoMoF_7$, $Na_2NiMnF_7$, $Na_2FeMoF_7$, $Na_2MnMoF_7$, $Na_2FeTiF_7$, and $Na_2Fe_2F_7$. In addition, a statistical analysis of equilibrium voltages for various redox couples demonstrated that all of these species exhibit higher redox potentials in fluoride chemistry when compared to polyanionic chemistry, such as phosphates. Fluorides, in comparison to other anionic chemistries, have received relatively little attention as potential electrode materials in current metal-ion battery technologies. Hence, the primary objective of this study is not only to showcase several promising cathode candidates but also to draw more attention to fluoride-based electrode materials in the battery community.

## DECLARATION OF INTERESTS

Miao Liu, Tenglong Lu, and Sheng Meng have a patent pending related to this work in the field of Na-ion battery.

## ACKNOWLEDGMENTS

This research is supported by National Key R&D Program of China (2021YFA0718700), Chinese Academy of Sciences (Grant No. CAS-WX2023SF-0101, ZDBS-LY-SLH007, and XDB33020000, XDB33030100, and YSBR-047), and the National Natural Science Foundation of China (Grand No. 12025407 and 11934003). The computational resource is provided by the Platform for Data-Driven Computational Materials Discovery of the Songshan Lake laboratory.

## REFERENCES

(1) Tian, Y.; Zeng, G.; Rutt, A.; Shi, T.; Kim, H.; Wang, J.; Koettgen, J.; Sun, Y.; Ouyang, B.; Chen, T.; Lun, Z.; Rong, Z.; Persson, K.; Ceder, G. Promises and Challenges of Next-Generation "Beyond Li-Ion" Batteries for Electric Vehicles and Grid Decarbonization. *Chem. Rev.* **2021**, *121* (3), 1623–1669. https://doi.org/10.1021/acs.chemrev.0c00767.

(2) Bianchini, M.; Wang, J.; Clément, R. J.; Ouyang, B.; Xiao, P.; Kitchaev, D.; Shi, T.; Zhang, Y.; Wang, Y.; Kim, H.; Zhang, M.; Bai, J.; Wang, F.; Sun, W.; Ceder, G. The Interplay between Thermodynamics and Kinetics in the Solid-State Synthesis of Layered Oxides. *Nat. Mater.* **2020**, *19* (10), 1088–1095. https://doi.org/10.1038/s41563-020-0688-6.

(3) Kim, H.; Ji, H.; Wang, J.; Ceder, G. Next-Generation Cathode Materials for Non-Aqueous Potassium-Ion Batteries. *Trends Chem.* **2019**,

*1* (7), 682–692. https://doi.org/10.1016/j.trechm.2019.04.007.

(4) Manthiram, A. A Reflection on Lithium-Ion Battery Cathode Chemistry. *Nat. Commun.* **2020**, *11* (1), 1550. https://doi.org/10.1038/s41467-020-15355-0.

(5) Barpanda, P.; Lander, L.; Nishimura, S.; Yamada, A. Polyanionic Insertion Materials for Sodium-Ion Batteries. *Adv. Energy Mater.* **2018**, *8* (17), 1703055. https://doi.org/10.1002/aenm.201703055.

(6) Shakoor, R. A.; Seo, D.-H.; Kim, H.; Park, Y.-U.; Kim, J.; Kim, S.-W.; Gwon, H.; Lee, S.; Kang, K. A Combined First Principles and Experimental Study on Na3V2(PO4)2F3 for Rechargeable Na Batteries. *J. Mater. Chem.* **2012**, *22* (38), 20535–20541. https://doi.org/10.1039/C2JM33862A.

(7) Saravanan, K.; Mason, C. W.; Rudola, A.; Wong, K. H.; Balaya, P. The First Report on Excellent Cycling Stability and Superior Rate Capability of Na3V2(PO4)3 for Sodium Ion Batteries. *Adv. Energy Mater.* **2013**, *3* (4), 444–450. https://doi.org/10.1002/aenm.201200803.

(8) Lemoine, K.; Hémon-Ribaud, A.; Leblanc, M.; Lhoste, J.; Tarascon, J.-M.; Maisonneuve, V. Fluorinated Materials as Positive Electrodes for Li- and Na-Ion Batteries. *Chem. Rev.* **2022**, *122* (18), 14405–14439. https://doi.org/10.1021/acs.chemrev.2c00247.

(9) Park, H.; Lee, Y.; Cho, M.; Kang, J.; Ko, W.; Jung, Y. H.; Jeon, T.-Y.; Hong, J.; Kim, H.; Myung, S.-T.; Kim, J. $Na_2Fe_2F_7$: A Fluoride-Based

Cathode for High Power and Long Life Na-Ion Batteries. *Energy Environ. Sci.* **2021**, *14* (3), 1469–1479. https://doi.org/10.1039/D0EE02803G.

(10)    Kang, J.; Ahn, J.; Park, H.; Ko, W.; Lee, Y.; Lee, S.; Lee, S.; Jung, S.-K.; Kim, J. Highly Stable Fe2+/Ti3+-Based Fluoride Cathode Enabling Low-Cost and High-Performance Na-Ion Batteries. *Adv. Funct. Mater.* **2022**, *32* (29), 2201816. https://doi.org/10.1002/adfm.202201816.

(11)    Liao, J.; Han, J.; Xu, J.; Du, Y.; Sun, Y.; Duan, L.; Zhou, X. Scalable Synthesis of Na$_2$MVF$_7$ (M = Mn, Fe, and Co) as High-Performance Cathode Materials for Sodium-Ion Batteries. *Chem. Commun.* **2021**, *57* (87), 11497–11500. https://doi.org/10.1039/D1CC04449D.

(12)    Kresse, G.; Joubert, D. From Ultrasoft Pseudopotentials to the Projector Augmented-Wave Method. *Phys. Rev. B* **1999**, *59* (3), 1758–1775. https://doi.org/10.1103/PhysRevB.59.1758.

(13)    Kresse, G.; Furthmüller, J. Efficiency of Ab-Initio Total Energy Calculations for Metals and Semiconductors Using a Plane-Wave Basis Set. *Comput. Mater. Sci.* **1996**, *6* (1), 15–50. https://doi.org/10.1016/0927-0256(96)00008-0.

(14)    Perdew, J. P.; Burke, K.; Ernzerhof, M. Generalized Gradient Approximation Made Simple. *Phys. Rev. Lett.* **1996**, *77* (18), 3865–3868. https://doi.org/10.1103/PhysRevLett.77.3865.

(15)    Dudarev, S. L.; Botton, G. A.; Savrasov, S. Y.; Humphreys, C. J.; Sutton, A. P. Electron-Energy-Loss Spectra and the Structural Stability of

Nickel Oxide: An LSDA+U Study. *Phys. Rev. B* **1998**, *57* (3), 1505–1509. https://doi.org/10.1103/PhysRevB.57.1505.

(16) Wang, L.; Maxisch, T.; Ceder, G. Oxidation Energies of Transition Metal Oxides within the GGA + U Framework. *Phys. Rev. B* **2006**, *73* (19), 195107. https://doi.org/10.1103/PhysRevB.73.195107.

(17) Jain, A.; Ong, S. P.; Hautier, G.; Chen, W.; Richards, W. D.; Dacek, S.; Cholia, S.; Gunter, D.; Skinner, D.; Ceder, G.; Persson, K. A. Commentary: The Materials Project: A Materials Genome Approach to Accelerating Materials Innovation. *APL Mater.* **2013**, *1* (1), 011002. https://doi.org/10.1063/1.4812323.

(18) Liang, Y.; Chen, M.; Wang, Y.; Jia, H.; Lu, T.; Xie, F.; Cai, G.; Wang, Z.; Meng, S.; Liu, M. A Universal Model for Accurately Predicting the Formation Energy of Inorganic Compounds. *Sci. China Mater.* **2023**, *66* (1), 343–351. https://doi.org/10.1007/s40843-022-2134-3.

(19) Lu, T.; Wang, Y.; Cai, G.; Jia, H.; Liu, X.; Zhang, C.; Meng, S.; Liu, M. Synthesizability of Transition-Metal Dichalcogenides: A Systematic First-Principles Evaluation. *Mater. Futur.* **2023**, *2* (1), 015001. https://doi.org/10.1088/2752-5724/acbe10.

(20) Hart, G. L. W.; Forcade, R. W. Algorithm for Generating Derivative Structures. *Phys. Rev. B* **2008**, *77* (22), 224115. https://doi.org/10.1103/PhysRevB.77.224115.

(21) Hart, G. L. W.; Nelson, L. J.; Forcade, R. W. Generating Derivative

Structures at a Fixed Concentration. *Comput. Mater. Sci.* **2012**, *59*, 101–107. https://doi.org/10.1016/j.commatsci.2012.02.015.

(22) Hart, G. L. W.; Forcade, R. W. Generating Derivative Structures from Multilattices: Algorithm and Application to Hcp Alloys. *Phys. Rev. B* **2009**, *80* (1), 014120. https://doi.org/10.1103/PhysRevB.80.014120.

(23) Morgan, W. S.; Hart, G. L. W.; Forcade, R. W. Generating Derivative Superstructures for Systems with High Configurational Freedom. *Comput. Mater. Sci.* **2017**, *136*, 144–149. https://doi.org/10.1016/j.commatsci.2017.04.015.

(24) Togo, A.; Tanaka, I. Spglib: A Software Library for Crystal Symmetry Search. arXiv August 5, 2018. https://doi.org/10.48550/arXiv.1808.01590.

(25) Toukmaji, A. Y.; Board, J. A. Ewald Summation Techniques in Perspective: A Survey. *Comput. Phys. Commun.* **1996**, *95* (2), 73–92. https://doi.org/10.1016/0010-4655(96)00016-1.

(26) Ong, S. P.; Richards, W. D.; Jain, A.; Hautier, G.; Kocher, M.; Cholia, S.; Gunter, D.; Chevrier, V. L.; Persson, K. A.; Ceder, G. Python Materials Genomics (Pymatgen): A Robust, Open-Source Python Library for Materials Analysis. *Comput. Mater. Sci.* **2013**, *68*, 314–319. https://doi.org/10.1016/j.commatsci.2012.10.028.

(27) Urban, A.; Seo, D.-H.; Ceder, G. Computational Understanding of Li-Ion Batteries. *Npj Comput. Mater.* **2016**, *2* (1), 16002.

https://doi.org/10.1038/npjcompumats.2016.2.

(28) Henkelman, G.; Uberuaga, B. P.; Jónsson, H. A Climbing Image Nudged Elastic Band Method for Finding Saddle Points and Minimum Energy Paths. *J. Chem. Phys.* **2000**, *113* (22), 9901–9904. https://doi.org/10.1063/1.1329672.

(29) Henkelman, G.; Jónsson, H. Improved Tangent Estimate in the Nudged Elastic Band Method for Finding Minimum Energy Paths and Saddle Points. *J. Chem. Phys.* **2000**, *113* (22), 9978–9985. https://doi.org/10.1063/1.1323224.

(30) Liu, M.; Rong, Z.; Malik, R.; Canepa, P.; Jain, A.; Ceder, G.; Persson, K. A. Spinel Compounds as Multivalent Battery Cathodes: A Systematic Evaluation Based on Ab Initio Calculations. *Energy Environ. Sci.* **2015**, *8* (3), 964–974. https://doi.org/10.1039/C4EE03389B.

(31) Foley, E. E.; Wu, V. C.; Jin, W.; Cui, W.; Yoshida, E.; Manche, A.; Clément, R. J. Polymorphism in Weberite $Na_2Fe_2F_7$ and Its Effects on Electrochemical Properties as a Na-Ion Cathode. *Chem. Mater.* **2023**, *35* (9), 3614–3627. https://doi.org/10.1021/acs.chemmater.3c00233.

(32) Jun, K.; Sun, Y.; Xiao, Y.; Zeng, Y.; Kim, R.; Kim, H.; Miara, L. J.; Im, D.; Wang, Y.; Ceder, G. Lithium Superionic Conductors with Corner-Sharing Frameworks. *Nat. Mater.* **2022**, *21* (8), 924–931. https://doi.org/10.1038/s41563-022-01222-4.

(33) Hautier, G.; Jain, A.; Ong, S. P.; Kang, B.; Moore, C.; Doe, R.;

Ceder, G. Phosphates as Lithium-Ion Battery Cathodes: An Evaluation Based on High-Throughput *Ab Initio* Calculations. *Chem. Mater.* **2011**, *23* (15), 3495–3508. https://doi.org/10.1021/cm200949v.

(34)     Ong, S. P.; Wang, L.; Kang, B.; Ceder, G. Li−Fe−P−O2 Phase Diagram from First Principles Calculations. *Chem. Mater.* **2008**, *20* (5), 1798–1807. https://doi.org/10.1021/cm702327g.

(35)     Ong, S. P.; Jain, A.; Hautier, G.; Kang, B.; Ceder, G. Thermal Stabilities of Delithiated Olivine MPO4 (M=Fe, Mn) Cathodes Investigated Using First Principles Calculations. *Electrochem. Commun.* **2010**, *12* (3), 427–430. https://doi.org/10.1016/j.elecom.2010.01.010.

(36)     Sun, W.; Dacek, S. T.; Ong, S. P.; Hautier, G.; Jain, A.; Richards, W. D.; Gamst, A. C.; Persson, K. A.; Ceder, G. The Thermodynamic Scale of Inorganic Crystalline Metastability. *Sci. Adv.* **2016**, *2* (11), e1600225. https://doi.org/10.1126/sciadv.1600225.

(37)     Ponrouch, A.; Marchante, E.; Courty, M.; Tarascon, J.-M.; Palacín, M. R. In Search of an Optimized Electrolyte for Na-Ion Batteries. *Energy Env. Sci* **2012**, *5* (9), 8572–8583. https://doi.org/10.1039/C2EE22258B.

(38)     Matts, I. L.; Dacek, S.; Pietrzak, T. K.; Malik, R.; Ceder, G. Explaining Performance-Limiting Mechanisms in Fluorophosphate Na-Ion Battery Cathodes through Inactive Transition-Metal Mixing and First-Principles Mobility Calculations. *Chem. Mater.* **2015**, *27* (17), 6008–6015. https://doi.org/10.1021/acs.chemmater.5b02299.

(39) Wang, X.; Tamaru, M.; Okubo, M.; Yamada, A. Electrode Properties of P2–Na$_{2/3}$Mn$_y$Co$_{1-y}$O$_2$ as Cathode Materials for Sodium-Ion Batteries. *J. Phys. Chem. C* **2013**, *117* (30), 15545–15551. https://doi.org/10.1021/jp406433z.

(40) Barpanda, P.; Oyama, G.; Nishimura, S.; Chung, S.-C.; Yamada, A. A 3.8-V Earth-Abundant Sodium Battery Electrode. *Nat. Commun.* **2014**, *5* (1), 4358. https://doi.org/10.1038/ncomms5358.

(41) Moreau, P.; Guyomard, D.; Gaubicher, J.; Boucher, F. Structure and Stability of Sodium Intercalated Phases in Olivine FePO4. *Chem. Mater.* **2010**, *22* (14), 4126–4128. https://doi.org/10.1021/cm101377h.

(42) Yuan, L.-X.; Wang, Z.-H.; Zhang, W.-X.; Hu, X.-L.; Chen, J.-T.; Huang, Y.-H.; Goodenough, J. B. Development and Challenges of LiFePO4 Cathode Material for Lithium-Ion Batteries. *Energy Env. Sci* **2011**, *4* (2), 269–284. https://doi.org/10.1039/C0EE00029A.

(43) Kawabe, Y.; Yabuuchi, N.; Kajiyama, M.; Fukuhara, N.; Inamasu, T.; Okuyama, R.; Nakai, I.; Komaba, S. Synthesis and Electrode Performance of Carbon Coated Na2FePO4F for Rechargeable Na Batteries. *Electrochem. Commun.* **2011**, *13* (11), 1225–1228. https://doi.org/10.1016/j.elecom.2011.08.038.